\documentstyle[epsfig]{mn}

\title[Non-Random Phases in Non-Trivial Topologies]{Non-Random Phases in Non-Trivial Topologies}

\author[P. Dineen, G. Rocha \& P. Coles]{Patrick
  Dineen$^{1}$\thanks{E-mail:ppxptd@nottingham.ac.uk}, Graca Rocha$^{2,3,4}$ \& Peter Coles$^{1}$\\
$^{1}$School of Physics \& Astronomy, University of Nottingham, University Park, Nottingham, NG7 2RD, United Kingdom\\
$^{2}$Astrophysics Group, Cavendish Laboratory, Madingley Road, Cambridge CB3 0HE, United Kingdom\\
$^{3}$Department of Physics, Nuclear \& Astrophysics Laboratory, University of Oxford, Keble Road, Oxford OX1 3RH, UK\\
$^{4}$Centro de Astrofisica da Universidade do Porto, R. das Estrelas s/n, 4150-762 Porto, Portugal}

\begin{document}
\maketitle

\begin{abstract}
We present a new technique for constraining the topology of the
universe. The method exploits the existence of correlations in the
phases of the spherical harmonic coefficients of the CMB
temperature pattern  associated with matched pairs of circles seen
in the sky in universes with non-trivial topology. The method is
computationally faster than all other statistics developed
to hunt for these matched circles. We applied the method to a range of
simulations with topologies of various forms and on different
scales. A characteristic form of phase correlation is found in the
simulations. We also applied the method to preliminary CMB maps derived from
WMAP, but the separation of topological effects from, e.g., foregrounds is not straightforward. 
\end{abstract}


\section{Introduction}
The 1st--year Wilkinson Microwave Anisotropy Probe (WMAP)
observations seem in good accord with the emerging standard
cosmological model; a flat $\Lambda$--dominated universe seeded by
scale-invariant adiabatic Gaussian fluctuations \cite{svpk03}.
However, at large scales there is an unexpected loss of CMB
anisotropy power that was also seen in the COBE-DMR data
\cite{bhhj03}. One possible explanation is that we inhabit a
universe that has a non--trivial topology. That is to say, our
universe is in fact multi-connected and has a finite volume. If
so, then power on scales exceeding the fundamental cell size will
be suppressed. Our space may not be large enough to support
long-wavelength fluctuations. A number of authors have tried to
restrict the topology of the universe using the WMAP data
\cite{lwrl03,cssk04,dtzh04,rlcm04,urlw04,wlrl04}. In particular,
Luminet et al. (2003) have shown that the Poincar\'e dodecahedral
space ($\Omega_o \approx 1.013$) accounts for the measured power
in the quadrupole and octopole better than an infinite
(simply--connected) flat universe. The angular power spectrum,
however, is not an effective way to characterise the peculiar form of
the anisotropy manifest in small universes of this type
\cite{lss98}.

Whether we can determine the topology of the universe depends on
its volume. If the universe is small enough, we should be able to
see right around it, since photons can travel across the whole
universe. If so, then we may be able to identify ghost images of
the same object in different directions in the sky or recognise
the topology from signatures in clustering statistics. Luminet \&
Roukema (1999) provide a review of these methods. However, these
methods are hindered by the need to identify good standard
candles; objects that can be traced through different eras. This
is not a problem when looking at the CMB for signatures of the
topology. The CMB photons originate from the same epoch and from a
very thin shell, the last scattering surface (LSS), which is the
same when viewed from either side. If the physical dimensions of
the universe are less than the diameter of the LSS then the sphere
self intersects; the loci of self-intersections are circles
\cite{css98}. We should therefore be able to match patterns of hot
and cold spots around circles. This result holds no matter how
complex the topology. A further advantage of using the CMB as an
indicator is that the last scattering surface marks the edge of
the visible universe, making it a powerful tool for looking for
non-trivial topology.

In this paper, we introduce a new method to search for evidence
 of a finite universe in all-sky CMB maps.
 Our method is based on the properties of the phases of the (complex)
 coefficients obtained from a spherical harmonic expansion of all-sky
 maps, specifically we look for phase correlations arising from matched pairs of circles.
  In the next section we briefly introduce some basic ideas in topology and discuss
  the simulations with non-trivial topologies used later on.
  In Section \ref{sec:phases} we sketch the procedure developed to detect
  phase correlations in the CMB. In Section \ref{sec:results} we discuss the results
  of applying the method to the simulations and WMAP data.
  We draw our conclusions in Section \ref{sec:conclusions}.

\section{Non-Trivial Topology}
\label{sec:topology} Topology deals with connectivity. To a
topologist, a coffee cup and a doughnut are equivalent, while a
coffee cup and a bowl are distinct. In the cosmological setting,
we are concerned with the connectivity of space: a manifold is
described as simply-connected if any closed path can be contracted
to a point. The possibility that our universe maybe
multi-connected was first suggested by Schwarzschild (1900), but
has often been overlooked in favour of the simplicity offered by
trivial topological spaces. Indeed there is a common misconception
that the shape of the universe can be determined from Einstein's
field equations, given a value of $\Omega_o$. General Relativity
only specifies the local curvature of space-time, so nothing in it
forbids a (global) non-trivial topology. The Cosmological
Principle merely restricts us to manifolds with constant
curvature. Indeed, any detection of non-trivial topology could
determine the sign of the spatial curvature since its
characteristics differ for the cases of Euclidean, spherical and
hyperbolic manifolds.

A 3-dimensional manifold can be described by identifying faces of
a fundamental cell/polyhedron. The WMAP data suggests that the
Universe is very nearly spatially flat, with $\Omega_o = 1.02 \pm
0.02$ \cite{bhhj03}. We therefore restrict our attention to
non-trivial topologies with a flat geometry. The technique should
nevertheless be applicable to spherical and hyperbolic spaces
where matched circles are expected.

In this study we use  6 compact orientable flat models that can be
constructed either by identifying the sides of a parallelepiped or
a hexagonal prism. The simplest one ({\bf Model 1}) is the
hypertorus (3-torus), which is obtained from a parallelepiped with
pairs of opposite faces glued together. This manifold is built out
of a parallelepiped by identifying $x\rightarrow x + h$,
$y\rightarrow y + b$ and $z\rightarrow z + c$. The next three
manifolds are variations of the hypertorus involving
identifications on opposite faces of a twisted parallelepiped. One
of these has opposite faces identified with one pair rotated
through $\pi$ (the twist torus; {\bf Model 2}). The next
identifies opposite faces with one rotated by $\pi/2$ (the $\pi/2$
twisted torus; {\bf Model 3}). The last of these three is obtained
by proceeding with the following identifications: ($x$, $y$,
$z$)$\rightarrow$($x + h$, $-y$, $-z$) corresponding to
translation along $x$ and rotation around $x$ by $\pi$; next ($x$,
$y$, $z$)$\rightarrow$($-x$, $y + b$, $-$($z +c$)) corresponding
to translation along $y$ and $z$ followed by rotation around $y$
by $\pi$; and finally ($x$, $y$, $z$)$\rightarrow$($-$($x + h$), $-$($y +
b$), $z + c$) translation along $x$, $y$ and $z$ followed by
rotation around $z$ by $\pi$  (the triple twist torus; {\bf Model
4}). Two other topologies are built out of a hexagon by
identifying the three pairs of opposite sides, while in the $z$
direction the faces are rotated relatively to each other by
$2\pi/3$ ($2\pi/3$ hexagon; {\bf Model 5}) or by
 $\pi/3$ ($\pi/3$ hexagon; {\bf Model 6}).

The simulations presented here are based on those in Rocha et al.
(2004): they have $\Omega_\Lambda=0$ and a Harrison--Zel'dovich
Gaussian spectrum. The simulations reflect topologies with
equal--sided physical dimensions that lead to a suppression of the
quadrupole with respect to the high--order modes \cite{wlrl04}. We
define a dimensionless  topological scale $j$ of a simulation as
the ratio between the width of the fundamental cell and the
horizon size.

These simulations include only those temperature fluctuations
generated by the Sachs--Wolfe effect. Matched circles occur
because we are looking at the same point on LSS from different
directions. In order for this to be true, the temperature
fluctuations need to be generated at the LSS which is true in this
case, but the SW effect only dominates at large scales. We
therefore  limit our investigation to $l \leq 20$. Even at these
scales, however, there are three main factors that could confuse
the statistic : (i) velocity perturbations generating anisotropies
at the LSS; (ii) the integrated Sachs-Wolfe effect due to time
varying potential wells crossed by the photons; and (iii) Galactic
foreground contamination of CMB data.

\section{Testing for Phase Correlations}
\label{sec:phases}
The temperature fluctuations in the CMB at any point in the celestial sphere
can be expressed in spherical harmonics as
\begin{equation}
\label{deltatovert} \Delta (\theta, \phi)=\sum _{l=1}^{\infty }\sum _
{m=-l}^{m=+l}a_{l,m}Y_{l,m}(\theta ,\phi ),
\end{equation}
where the $a_{l,m}$ are complex coefficients that can be written
\begin{equation}
a_{l,m}=|a_{l,m}|\exp[i\phi_{l,m}].
\end{equation}
In orthodox cosmologies the temperature fluctuations constitute a
statistically homogeneous and isotropic Gaussian random field. In
this case the phases $\phi_{l,m}$ are independent and uniformly
random on the interval [0,2$\pi$] and the variance of $a_{lm}$
depends only upon $l$. Departures from orthodoxy lead to
differences in behaviour of the $a_{l,m}$. For example, in
anisotropic Gaussian fluctuations the variance of the $a_{l,m}$
depends on $m$ \cite{fm97,is03}.  Coles et al. (2004) developed a
diagnostic of departures from the standard assumption that
involves the randomness of phases. The main component of the
technique involved using Kuiper's statistic from an available set
of phase angles. First the phases are sorted into ascending order,
to give the set $\{\theta _{1},\ldots ,\theta _{n}\}$. Each angle
$\theta _{i}$ is divided by $2\pi$ to give a set of variables
$X_{i}$, where $i=1\ldots n$. From the set of $X_i$ we derive two
values $S^+_{n}$ and $S^-_{n}$ where
\begin{equation}
S^{+}_{n} = {\rm max}
\left\{\frac{1}{n}-X_{1},\frac{2}{n}-X_{2},\ldots ,1-X_{n}\right\}
\end{equation}
and   \begin{equation} S^{-}_{n} = {\rm max}
\left\{X_{1},X_{2}-\frac{1}{n},\ldots
,X_{n}-\frac{n-1}{n}\right\}.
\end{equation}
Kuiper's statistic, $V$, is then defined as
\begin{equation}
\label{TestStatisticV} V=(S^{+}_{n}+S^{-}_{n})\cdot
\left(\sqrt{n}+0.155+\frac{0.24}{\sqrt{n}}\right).
\end{equation}
The form of $V$ is chosen so that it is approximately independent of sample size for large $n$.
Anomalously large values of $V$ indicate a distribution that is more clumped than a uniformly random distribution,
while low values mean that angles are more regular. In order to remove any artifact from the
choice of coordinate frame that the $a_{l,m}$ are measured in, $V$ is calculated for randomly rotated coordinate systems.
The rotated coefficients are found by employing the Wigner $D$ function
\begin{equation}
\label{RotatedCoefficients} a_{l,m}=\sum _{m^{\prime
}}a_{l,m^{\prime }}D^{l}_{m,m^{\prime }}(\alpha ,\beta ,\gamma ).
\end{equation}
where the Euler angles $\alpha$, $\beta$, $\gamma$ define the
magnitude of successive rotations about the coordinate axes.
Consequently, a distribution of $V$ is obtained from one set of
phases. This distribution is compared, using a $\chi^2$ test, with
distributions obtained in a similar manner from 1,000 Monte Carlo
(MC) sets of $\phi_{l,m}$ drawn from a uniformly random
distribution. The probability $\cal P$($\chi^2$) of obtaining a
lower value of $\chi^2$ is calculated from the fraction of
$\chi^2_{\mathrm{MC}}$ that are less than the $\chi^2$ obtained from
the phases. A set of angles is assumed to be non-random if $\cal
P$($\chi^2$) $\geq$ 0.95.

The random-phase hypothesis can be further scrutinised by investigating subsets of the phases.
These subsets should also be uniformly random on the interval [0,2$\pi$].
In this paper, we look at two subsets: (i) the phase differences for fixed values of
$m$ ($\phi_{l+1,m}-\phi_{l,m}$) and (ii) the phase differences for fixed values of $l$ ($\phi_{l,m+1}-\phi_{l,m}$)
(even $l$-modes only). Both subsets are of particular interest since Cornish et al. (2004)
indicate that matched circles are associated with phase correlations.
In their paper, the significance level for detection of matched circles in the WMAP data is
calculated from 'scrambled' versions of the data. In the scrambled versions, phase correlations are
removed by randomly exchanging the $a_{l,m}$ at fixed $l$. Also, as previously mentioned multi-connectedness
breaks the global isotropy and sometimes the global homogeneity of the Universe (except the case of the projective space).
 This will induce correlations between the $a_{l,m}$ of different $l$ and $m$.
 For instance, due to the symmetries of the hypertorus case,
$\langle a_{l,m} a^{*}_{l',m'} \rangle \neq 0  \Rightarrow m-m' \equiv 0
 \bmod(2)$ and $ l -l' \equiv 0 \bmod(2)$ \cite{rulw04}.
 These non-zero off-diagonal terms will induce phase correlations as detected in our study.

Overall, for a given temperature map, we obtain a value of $\cal
P$($\chi^2$) for each mode; 18 values for subset (i) and 10 for
subset (ii). To improve the presentation of the results, we
combine $\cal P$($\chi^2$) for each subset in two ways. First, we
count the number of modes with ${\cal P}(\chi^2) \geq 0.95$
 and find the mode that displays the highest value. Secondly, we
 perform a Kolmogorov-Smirnov (K-S) test on the distribution
of $\cal P$($\chi^2$) over the modes. If the phases are random,
the set of $\cal P$($\chi^2$) should be uniform in the interval
$[0,1]$. To quantify the significance of the K-S value obtained,
10,000 MC sets of $\cal P$($\chi^2$) are generated and the
probability of obtaining a lower K-S value is calculated. The
reason for doing both these things is that one would expect one in
twenty modes to yield a value of ${\cal P}(\chi^2) \geq 0.95$. The
second approach gives a (very conservative) idea of the
significance of the whole set of modes rather than each individual
one.

\section{Results and Discussion}
\label{sec:results}

The simulations were generated in HEALPix\footnote{http://www.eso.org/science/healpix/} format \cite{healpix}
with a resolution parameter $N_{\mathrm{side}}$=$32$.
The $a_{l,m}$ were derived using the '{\tt anafast}' routine in the HEALPix
package. $V$ was binned from $0$--$2.75$ and $0$--$2.5$ for subsets (i) and (ii),
 respectively.
Subset (i) required 10,000 rotations in order to obtain stable results.
To analyse one realisation took 18 minutes on 1,400 MHz CPU desktop.
On the other hand, 3,000 rotations produced stable results for subset (ii),
 resulting in each analysis taking $2\frac{3}{4}$ minutes on the same
 desktop.

The six flat models listed in Section \ref{sec:topology}
were studied with $j$=$0.5$. In order to see the effect $j$
had on the results, the hypertorus was explored in more detail.
 A further six simulations with $j<2$ were scrutinised.
 Also, six simulations in which matched circles are not anticipated ($j \geq 2$) were studied.
 The results for subsets (i) and (ii) are shown in Tables \ref{tab:D_l(m)} and \ref{tab:D_m(l)} respectively.
 The tables show the number of modes with $\cal P$($\chi^2$) $\geq$ 0.95,
 the most non--random mode and the K--S fractional probabilities. For subset
 (i), the method was applied to five realisations (sets of rotations)
 of each simulation. The average values obtained from these realisations are
 shown in the tables. For subset (ii), only one realisation was necessary in
 order to obtain consistent results.

If the phases are random, the number of modes with $\cal
P$($\chi^2$) $\geq$ $0.95$ should be $0.9$ for subset (i) and $0.5$ for
subset (ii). From Table \ref{tab:D_l(m)} it is clear that we are
finding correlations when scanning across fixed $m$ for most of the
simulations (we shall refer to these
as $m$--correlations). The $m$--correlations are less obvious in
terms of the K--S probabilities. However, this test is more
general: it does not search specifically for large values of $\cal
P$($\chi^2$), so it should be interpreted as a very crude measure
of departure from uniformity. The average count for all the simulations displayed in the
Table \ref{tab:D_l(m)} is $2.5$. The average value is significantly larger than the expected value of
$0.9$. This contrasts with the average value of $0.7$, from Table \ref{tab:D_m(l)}, that is only slightly
higher than the expected value. 
From the results, it is evident that no particular mode can be chosen to look for signs of non--randomness in the phases.
This is unsurprising, as the correlations would manifest themselves across
many modes, whose nature would depend on the location and size of the circles with respect to the observer.

The spread in the number count from one realisation to another is quite small
(roughly $\pm1$ for larger values) and an exact value for each simulation can be
obtained by increasing the number of realisations.
To see if there is any worth in doing this, we looked at five further simulations for each model with $j$=$0.5$,
the results of which are shown in Table \ref{tab:D_l(m)2}. The standard
deviation is very large, with a count of $1.0$ being consistent with all the models.
 This indicates that it would be very difficult to use the method to distinguish between models or to
 determine the exact topological scale, you could merely indicate the most probable case.
 These results confirm that the hypertorus and the $\pi/3$ hexagon models,
 at this scale, do not display any $m$--correlations in the phases. However,
 the triple twist torus is detectable from $m$--correlations. This fact is
 less clear in Table \ref{tab:D_l(m)}, again indicating that the test
 results vary from simulation to simulation. 

Looking at results for the hypertorus displayed in Table \ref{tab:D_l(m)},
it is hard to perceive any trend between the number count and the
topological scale of the simulation. The $m$--correlations are
seen both with values of $j\!<\!2$ and $j\!\geq\!2$. Intuitively, one would
expect phase correlations when matched circles are present ($j\!<\!2$). In
such cases, the number of repeated patterns (circles in the sky) increases with decreasing $j$, and
hence we would expect an increase in number count with decreasing $j$. This is not
seen in the results, although, this may be masked by the number count varying
from simulation to simulation for fixed values of $j$. Applying the statistic
to further simulations may reveal such a trend.

On the other hand, the high number count seen at $j\!\geq\!2$ is less easy to
explain. Unearthing signs of non--trivial topology beyond the horizon size,
suggests the technique is potentially a powerful tool. Diagnostics that hunt for circles in the sky are limited to
a maximum value of $j$=$2$. Our detections beyond the horizon size are
by no means serendipitous. Phillips \& Kogut (2004) compute the covariance matrix of the
$a_{l,m}$ for the hypertorus at various topological scales. They find the
off--diagonal terms, that incorporate the phase information, remain prominent
even when the width of the fundamental cell is greater than the diameter of the LSS.

Apart from the hypertorus, each of the topologies addressed are not only
anisotropic, but also inhomogeneous. The question therefore arises whether the
results are affected greatly by changes in the position of the observer. We
generated simulations with the observer's position shifted within the
fundamental cell. For models 2 to 6, five simulations with $j$=$0.5$ were
generated with the observer position shifted by ($0.1x$,$0.1x$,$0.1x$) where $x$
is the width of the fundamental cell. The number count for each model was
found to be consistent with those of the
centrally located observer displayed in Table \ref{tab:D_l(m)2}.

A positive detection of $m$-correlations in CMB data would
therefore be indicative of the universe having a non-trivial
topology. In order to seek evidence of these correlations in the
CMB data, we turned to four WMAP-derived maps. The temperature
maps were all constructed in a manner that minimises foreground
contamination and detector noise, leaving a pure CMB signal. The
ultimate goal of these maps is to build an accurate image of the
LSS that captures the detailed morphology. Following the release
of the WMAP 1 yr data, the
 WMAP team \cite{bhhn03}, Tegmark, de Oliveira-Costa \& Hamilton (2003), Naselsky et al. (2003) and
  Eriksen et al. (2004a) have released CMB-only sky maps (see papers for details).
  We shall refer to these as the ILC, TOH, Naselsky and Eriksen maps respectively.
  It is worth pointing out that even though we limit ourselves to full sky maps in this paper,
  the method can be adapted to cut maps following a suitable adjustment of the MC simulations. Nevertheless, the
  current WMAP data is preliminary, so we reserve this treatment for future
  data releases. Again, the results are
shown in Tables \ref{tab:D_l(m)} and \ref{tab:D_m(l)}. The method was
applied to five realisations of each CMB map for subset (i).

In three of the four maps, evidence was found for the sort of phase
correlations seen in the simulations. However, all the maps, bar the Naselsky map, display $l$--correlations that
were discussed in Coles et al. (2004) which can be explained, at
least partially, by foreground contamination. If no $m$--correlations were seen
then this would suggest there is no evidence for non--trivial
topology. However, a positive detection leaves open the possibility of a
non--trivial topology. 
  
\begin{table}\centering
\begin{tabular}{l|l|c}\hline\hline
\emph{Map} & \emph{Count} (\emph{Mode})&\emph{K-S Probability} \\
\hline
Model 1 $j$=0.5 &0.6 & 0.68\\
Model 2 $j$=0.5 &1.2 (17)& 0.35\\
Model 3 $j$=0.5 &4.6 (4,5)& 0.99\\
Model 4 $j$=0.5 &0.6 & 0.22\\
Model 5 $j$=0.5 &1.8 (14)& 0.68\\
Model 6 $j$=0.5 &0.2 & 0.42\\
\hline
Model 1 $j$=0.4 & 5.4 (1,4,9,12)& 0.99\\
Model 1 $j$=0.6 & 0.4& 0.63\\
Model 1 $j$=0.8 & 0.8& 0.35\\
Model 1 $j$=0.9 & 0.0& 0.56\\
Model 1 $j$=1.0 & 0.4& 0.47\\
Model 1 $j$=1.6 & 3.8 (8,9,10)& 0.77\\
Model 1 $j$=2.0 & 1.4 (16) & 0.43\\
Model 1 $j$=4.0 & 0.0& 0.66\\
Model 1 $j$=5.0 & 2.8 (12)& 0.48\\
Model 1 $j$=6.0 & 0.0& 0.62\\
Model 1 $j$=8.0 & 0.6& 0.23\\
Model 1 $j$=10.0& 2.1 (12) &0.88\\
\hline
ILC & 1.2 (4)& 0.55\\
TOH & 0.6 & 0.28\\
Naselsky & 2.2 (2,4)& 0.50\\
Eriksen & 2.0 (1,4)& 0.33\\
\hline\hline
\end{tabular}
\caption{ \label{tab:D_l(m)} Phase correlations when scanning
across fixed $m$. The column labelled \emph{Count} shows the
average number of modes exceeding 95 per cent significance. The
column labelled \emph{Mode} gives the mode with greatest
significance. The last column shows a rough measure of
significance for all modes obtained using a K-S test as described
in the text. 18 modes studied in total.}
\end{table}

\begin{table}\centering
\begin{tabular}{l|l|c}\hline\hline
\emph{Map} & \emph{Count} (\emph{Mode})&\emph{K-S Probability} \\
\hline
Model 1 $j$=0.5 & 0& 0.17\\
Model 2 $j$=0.5 & 1& 0.66\\
Model 3 $j$=0.5 & 0& 0.52\\
Model 4 $j$=0.5 & 1& 0.45\\
Model 5 $j$=0.5 & 0& 0.46\\
Model 6 $j$=0.5 & 2 (2,12)& 0.13\\
\hline
Model 1 $j$=0.4 & 2 (8,20) & 0.32\\
Model 1 $j$=0.6 & 0& 0.21\\
Model 1 $j$=0.8 & 0& 0.68\\
Model 1 $j$=0.9 & 1& 0.97\\
Model 1 $j$=1.0 & 1 & 0.08\\
Model 1 $j$=1.6 & 0&  0.09\\
Model 1 $j$=2.0 & 1&  0.49\\
Model 1 $j$=4.0 & 1&  0.47\\
Model 1 $j$=5.0  & 1&  0.16\\
Model 1 $j$=6.0 & 1&  0.54\\
Model 1 $j$=8.0 & 1&  0.73\\
Model 1 $j$=10.0 & 0&  0.43\\
\hline
ILC & 2 (14,16)& 0.50\\
TOH      & 3 (6,14,16)&  0.77\\
Naselsky & 1 (14)&  0.81\\
Eriksen  & 2 (6,16)&  0.80\\
\hline\hline
\end{tabular}
\caption{ \label{tab:D_m(l)} Phase correlations when scanning
across fixed $l$. Columns are as in the previous table. 10 even
modes studied in total.}
\end{table}

\begin{table}\centering
\begin{tabular}{l|l}\hline\hline
\emph{Map} & \emph{Count} (\emph{Mode})\\
\hline
Model 1 & 0.5 $\pm$ 0.2\\
Model 2 & 1.0 $\pm$ 0.9\\
Model 3 & 3.4 $\pm$ 3.4\\
Model 4 & 1.0 $\pm$ 0.7\\
Model 5 & 1.3 $\pm$ 0.8\\
Model 6 & 0.8 $\pm$ 0.5\\
\hline\hline
\end{tabular}
\caption{ \label{tab:D_l(m)2} The variation in number count from
simulation to simulation. The columns show the average number of
modes (along with the variation) and the most significant mode. We
evaluated 6 simulations with $j$=0.5  for each model.}
\end{table}

\section{Conclusions}
\label{sec:conclusions} In this paper, we have presented a new
method for constraining the topology of the universe. The method
relies on utilising phase correlations associated with matched
pairs of circles in the CMB sky. We applied the method to various
simulations with non-trivial topologies. 
The method appears to detect non--trivial topologies beyond the horizon size.
However, the method fails to
estimate the scale of
features it detects and is not good at discriminating between
different models. This can potentially be overcome by studying more
simulations as the results were shown to vary from simulation to simulation.
The method is simple,
computationally fast and does deliver a clear signature:
 a positive detection of these $m$-correlations is
clear evidence for non-trivial topology.

With this in mind, the method was applied to four CMB--only sky
maps; we found evidence for $m$--correlations in three of them.
However, it would be premature to conclude that there is evidence for
non--trivial topology in any of the available CMB temperature map. The WMAP data
is preliminary and has already been shown to have a number of
unusual properties that are not yet fully understood
\cite{cnvw03,dc04,ehbg04}. Indeed, Eriksen et al. (2004a) have
pointed out that the techniques used in producing these maps
result in a poor reconstruction of the cosmological phases which may interfere
with the possibility of detecting correlations of the type discussed
here. Nevertheless, with improved data, we believe the method will be a
useful tool in determining the shape of the universe. Moreover,
our method is based on only the simplest possible measure of
randomness in the phase distribution. More sophisticated
combinations may allow us to improve the method substantially,
perhaps unearthing further signs of non--trivial topology beyond the
horizon size.

\section*{Acknowledgements}
GR would like to thank Janna Levin, Evan Scannapieco and Lance Miller for
fundamental comments on the simulations of non-trivial topologies. GR would
like to acknowledge a Leverhulme fellowship at the University of Cambridge.

\end{document}